\documentclass[paper,usegraphicx,useAMS,usenatbib]{mn2e}
\usepackage{epsfig}
\input{epsf}

\def\beq{\begin{equation}}
\def\ee{\end{equation}}
\def\lsim{\mathrel{\rlap{\lower4pt\hbox{\hskip1pt$\sim$}}
    \raise1pt\hbox{$<$}}}
\def\gsim{\mathrel{\rlap{\lower4pt\hbox{\hskip1pt$\sim$}}
    \raise1pt\hbox{$>$}}}

\def\ts{\times}

\def\kms{\rm km\:s^{-1}}

\def\lesssim{\mathrel{\hbox{\rlap{\hbox{\lower4pt\hbox{$\sim$}}}\hbox{$<$}}}}
\def\gtrsim{\mathrel{\hbox{\rlap{\hbox{\lower4pt\hbox{$\sim$}}}\hbox{$>$}}}}

\title[Star-Forming Accretion Flows]
{Star-forming accretion flows and the low luminosity
nuclei of giant elliptical galaxies}
\author[J. C. Tan \& E. G. Blackman] 
{Jonathan C. Tan$^{1,2}$ and Eric G. Blackman$^{3}$\\
$^1$Institute of Astronomy, Department of Physics, ETH Z\"urich, 8093 Z\"urich, Switzerland\\
$^2$Department of Astrophysical Sciences, Peyton Hall,
Princeton Universtiy, Princeton, NJ 08544, USA\\
$^3$Department of Physics and Astronomy, University of Rochester, 
Rochester, NY 14627, USA}

\begin{document}

%\date{Accepted. Received; in original form}

\pagerange{\pageref{firstpage}--\pageref{lastpage}} \pubyear{2005}

\maketitle

\label{firstpage}

\begin{abstract} 
  The luminosities of the centers of nearby elliptical galaxies are
  very low compared to models of thin disc accretion to their black
  holes at the Bondi rate, typically a few hundredths to a few tenths
  of a solar mass per year. This has motivated models of
  inefficiently-radiated accretion that invoke weak electron-ion
  thermal coupling, and/or inhibited accretion rates due to convection
  or outflows.  Here we point out that even if such processes are
  operating, a significant fraction of the accreting gas is prevented
  from reaching the central black hole because it condenses into stars
  in a gravitationally unstable disc. Star formation occurs inside the
  Bondi radius (typically $\sim 100$~pc in giant ellipticals), but
  still relatively far from the black hole in terms of Schwarzschild
  radii. Star formation depletes and heats the gas disc, eventually
  leading to a marginally stable, but much reduced, accretion flow to
  the black hole. We predict the presence of cold ($\sim 100\:{\rm
  K}$), dusty gas discs, containing clustered H$\alpha$ emission and
  occasional type II supernovae, both resulting from the presence of
  massive stars.  Star formation accounts for several features of the
  M87 system: a thin disc, traced by H$\alpha$ emission, is observed
  on scales of about 100~pc, with features reminiscent of spiral arms
  and dust lanes; the star formation rate inferred from the intensity
  of H$\alpha$ emission is consistent with the Bondi
  accretion rate of the system.  Star formation may therefore help
  suppress accretion onto the central engines of massive
  ellipticals. We also discuss some implications for the fueling of
  the Galactic center and quasars.
\end{abstract}

\begin{keywords}
accretion: accretion discs -- black hole physics -- galaxies: active -- galaxies: individual(M87) -- galaxies: elliptical and lenticular -- stars: formation
\end{keywords}

\section[]{Introduction}\label{sec1} 

Optical spectroscopy and photometry provide good evidence that black
holes with masses of $10^8-10^{10}M_\odot$ reside at the centers of
giant elliptical galaxies (e.g. Kormendy \& Richstone 1995).  X-ray
observations of these galaxies allow an inference of the interstellar
gas temperature and thus an estimate of the Bondi accretion radius,
where the dynamics of the gas start to be dominated by the potential
of the black hole. The density of the gas is also derived from these
observations, so the Bondi accretion rate can be estimated. The high
resolution of the Chandra X-ray Observatory has proved to be
particularly useful for such studies.

The bolometric luminosities of the black hole engines can be estimated
from multi-wavelength observations, particularly in the infrared.
These luminosities are typically several orders of magnitude fainter
than predicted by models of thin disc accretion at the Bondi rate in
which about 10\% of the rest mass energy is radiated (e.g. Fabian \&
Rees 1995). A particularly well-studied case is M87 (3C~274; NGC~4486)
by Di Matteo et al. (2000, 2003) (see also Reynolds et al. 1996a), but
similar results have also been found for NGC 1399, NGC 4472, and NGC
4636 (M49) (Di Matteo et al. 2000; Loewenstein et al. 2001), for NGC
4697 (Di Matteo et al. 2000; Sarazin, Irwin \& Bregman 2001) for NGC
4697, and for NGC 4649 (M60) (Di Matteo et al. 2000; Randall, Sarazin
\& Irwin 2004).

In order to explain such quiescent accretors, two temperature
collisionless accretion models have been invoked (e.g., Ichimaru 1977;
Narayan \& Yi 1995; Quataert \& Gruzinov 1999; 
%Narayan, Igumenshchev \& Abramowicz 2000; 
Narayan 2002).  These models typically assume that the energy exchange
between electrons and ions proceeds by Coulomb collisions. The
associated accretion flows do not radiate efficiently if (1) the
dissipation primarily heats the protons and (2) if the electron-ion
equilibration time is is longer than the radial infall time. The
gravitational energy is then carried by ions to the event horizon of
the black hole, in what are known as Advection-Dominated Accretion
Flows (ADAFs).

For the ellipticals, minimalistic standard ADAF models predict too
high a luminosity (e.g. Di Matteo et al. 2000). Variations of ADAFs
are however still in the running: ADAFs can be combined or replaced 
with a models involving a 
radially dependent accretion rate and winds (Blandford \& Begelman 1999;
Quataert \& Narayan 2000). Alternatively, ADAFs unstable to convection
(called convection dominated accretion flows -- CDAFs; Narayan,
Igumenshchev, \& Abramowicz 2000; Quataert \& Gruzinov 2000) have the
outward transport of angular momentum inhibited by the convection and
thus their accretion rate onto the central engine reduced.

There are however some unresolved issues with the physics of these
models. Regarding CDAFs, the parameter space allowed for solutions
that reduce the outward transport of angular momentum becomes
restricted when magnetic fields are present and seems to require that
the field saturates at a very low value.  Otherwise the
magneto-rotational instability dominates the outward angular momentum
transport (Balbus \& Hawley 2002; Narayan et al. 2002).  It should
also be noted that virtually all flavors of optically thin ADAFs and
CDAFs require maintenance of two-temperature collisionless flows.  The
unanswered plasma physics questions about the viability of this
situation have also been an important area of ongoing research
(Begelman \& Chiueh 1988; Bitsnovatyi-Kogan \& Lovelace 1997, Quataert
1998, Gruzinov 1998, Blackman 1999; Quataert \& Gruzinov 2000;
Blackman 2000; Pariev \& Blackman 2003).

For these reasons it is still of interest to consider other
contributors to the quiescence of elliptical galaxy centers. Here
we propose a very simple alternative (or companion process) 
to these accretion models.  We 
consider the possibility that most of the accreting gas forms stars,
shutting off the supply of fuel to the engine.  In \S2 we make a careful
estimate of the Bondi accretion rate in elliptical galaxies. In \S3 we
then describe the expected properties of the disc that forms inside
the Bondi radius, concluding that it is gravitationally unstable and
should form stars. In \S4 we consider general properties of
star-forming ``accretion'' discs around black holes, describing the
signatures and feedback from star formation, its efficiency, and the
amount of gas present in the disc in steady state.  We discuss the
application of this model to M87, other ellipticals, the Galactic
center, quasars and AGN in \S5, and conclude in \S6.

\section{The Accretion Rate\label{S:accrete}}

X-ray observations show that elliptical galaxies contain hot gas
approximately at the virial temperature (e.g. Mathews \& Brighenti
2003). Where high resolution data are available, there is often
evidence for a temperature gradient such that the central temperature
is cooler than the outer regions (e.g. M87 --- Di Matteo et al. 2003;
NGC~4472, NGC~4636 and NGC~5044 --- Matthews \& Brighenti 2003 and
references therein). Typical temperatures on scales of several hundred
parsecs from the galaxy centers are $kT\sim 1$~keV and so the gas is
fully ionized with a mean particle mass of $\mu m_{\rm H}$ with $\mu
\simeq 0.6 $. We define $\mu_{0.6} \equiv \mu/0.6$. This gas has an
effective isothermal sound speed of $c_s\equiv (P/\rho)^{1/2}= 400
\mu_{0.6}^{-1/2} (kT_\beta/{\rm keV})^{1/2}\kms$, where the total
pressure is $P\equiv nkT_\beta$ and $\rho$ is the density.  It is
possible that thermal pressure accounts for only a fraction, $\beta$,
of the total pressure (the remainder may include magnetic, turbulent
and cosmic ray pressures) and we account for this by defining
$T_\beta\equiv T/\beta$ to be an effective temperature that is a
factor $\beta^{-1}$ higher than the actual gas temperature, $T$, i.e.
as inferred from X-ray observations.  However, to simplify the
equations, henceforth we shall assume $\beta=1$ and use $T_{\rm keV}$
in place of $k T/(\beta {\rm keV})$. To allow for nonthermal pressure,
one can simply substitute $T_{\rm keV}\rightarrow T_{\rm keV}/\beta$.

An object of mass $M$ begins to have a significant gravitational
influence on the gas at a distance $r_{\rm B}$, the Bondi radius. Note
that in the case of supermassive black holes in giant elliptical
galaxies, $M$ may also include a contribution from a concentrated
stellar cluster around the black hole. Following the notation of Shu
(1992), the Bondi radius is 
\beq r_{\rm B}\equiv \frac{GM}{c_{s,\infty}^2}=26.9 \mu_{0.6,\infty} \frac{M_9}{T_{\rm keV, \infty}} \:{\rm pc},
\label{rbondi}
\ee where $M_9=M/10^9M_\odot$ and subscript $\infty$ refers to a
location far from the black hole, i.e. at least several $r_{\rm B}$.

We can express the actual accretion rate as being some fraction
$f_{\rm B}$ of the Bondi rate 
\beq
\begin{array}{l}
\dot{M} =  f_{\rm B} \dot{M}_{\rm B}
= f_{\rm B} \lambda_\gamma 4 \pi \rho_\infty (GM)^2 / c_{s,\infty}^3,\\
  =  0.0150 f_{\rm B} \frac{\lambda_\gamma}{0.272} n_\infty \mu_{0.6,\infty}^{5/2} M_9^2 T_{\rm keV, \infty}^{-3/2}\:M_\odot\:{\rm yr^{-1}},
\end{array}
\label{mdot}
\ee where $\lambda_\gamma$ is a dimensionless factor that depends on
the effective equation of state of the gas near the Bondi radius
($\lambda_\gamma=0.116,0.272,0.377,1.12$ for $\gamma=5/3,3/2,7/5,1$;
Shu 1992, p437). The factor $f_{\rm B}$ takes account of effects such
as the inhibition of accretion from certain directions because of an
outflow from the galactic central engine.

The expected luminosity of the nuclear region inside $r_{\rm B}$ can be
expressed as
\beq
L = \epsilon \dot{M} c^2 = 8.50 \times 10^{43} \epsilon_{0.1} f_{\rm B} \frac{\lambda_\gamma}{0.272} \mu_{0.6,\infty}^{5/2} \frac{n_\infty M_9^2}{T_{\rm keV, \infty}^{3/2}}\:{\rm ergs\:s^{-1}},
\label{lbondi}
\ee where $\epsilon_{0.1}\equiv\epsilon/0.1$, with this canonical
value being appropriate for thin disc accretion.  The fact that
luminosities of the nuclear regions of many giant ellipticals are less
than the above value has motivated models of either inefficient
radiation (small $\epsilon$) or inefficient accretion (small $f_{\rm
  B}$).  We can relate this luminosity to the Eddington value:
\beq \frac{L}{L_{\rm E}} = 6.81 \times 10^{-4} \epsilon_{0.1} f_{\rm B}
\frac{\lambda_\gamma}{0.272} \mu_{0.6,\infty}^{5/2} \frac{n_\infty
  M_9}{T_{\rm keV, \infty}^{3/2}}.
\label{lbondiedd}
\ee

Now consider the heating and cooling rates at the Bondi radius. Their
relative values determine which value of $\gamma$ is most appropriate
in the equation of state. We assume the heating is only due to
compression. The heating rate per unit volume of gas in a spherical
shell at $r_{\rm B}$ is \beq
\begin{array}{r}
\Gamma = 2 P_{\rm B} u_{r,{\rm B}} / r_{\rm B} = 2 f_\rho^\gamma n_\infty k T_\infty \beta_{\rm B}^{-1} f_u c_{s,\infty} /r_{\rm B}\\
= 1.54\ts10^{-21} f_\rho^\gamma f_u \mu_{0.6,\infty}^{-3/2} T_{\rm keV,\infty}^{5/2}  n_\infty M_9^{-1} \:{\rm ergs\:cm^{-3}\:s^{-1}},
\end{array}
\label{heat}
\ee where subscript $\rm B$ refers to the Bondi radius, $f_\rho \equiv
\rho_{\rm B}/\rho_\infty = 1.44,1.55,1.64,2.45$ for
$\gamma=5/3,3/2,7/5,1$ and $f_u\equiv u_{r,{\rm
    B}}/c_{s,\infty}=0.78,0.72,0.69,0.46$ for $\gamma=5/3,3/2,7/5,1$, and
where in the last expression we have assumed $\beta_{\rm
  B}=\beta_\infty = 1$.  The cooling rate for gas under conditions
typical of the centers of ellipticals has been calculated by
Sutherland \& Dopita (1993).  For $0.0274\:{\rm keV}<kT<2.74$~keV and
for metallicities that are about a factor of three greater than solar,
the cooling is dominated by resonance line emission from O, Ne and Fe
and can be approximated as \beq \Lambda = 6.38\times 10^{-23}
\frac{Z}{3.2Z_\odot} T_{\rm keV}^{-0.71} n^2\:{\rm ergs
  \:cm^{-3}\:s^{-1}}.
\label{cool}
\ee
We can use equations (\ref{heat}) and (\ref{cool}) to evaluate when
heating balances cooling at the Bondi radius, so that the isothermal
$\gamma=1$ solution is most applicable. This condition is 
\beq
n_\infty M_9 = 24.1 f_\rho^{\gamma-2+0.71(\gamma-1)} f_u T_{\rm keV,\infty}^{3.21}\:{\rm cm^{-3}},
\label{balance}
\ee which is also shown in Figure 1. If at a given temperature the
combination $n_\infty M_9$ is less than this isothermal condition,
then  equations of state with $1<\gamma<5/3$ are more
appropriate, as they account for the inability of the
gas to cool fast enough to maintain the isothermal state. 
Note that while we have included the $\beta$ dependence
in the heating rate, we have not allowed for any ``cooling'' of the
nonthermal component. 

\begin{figure}
\epsfig{file=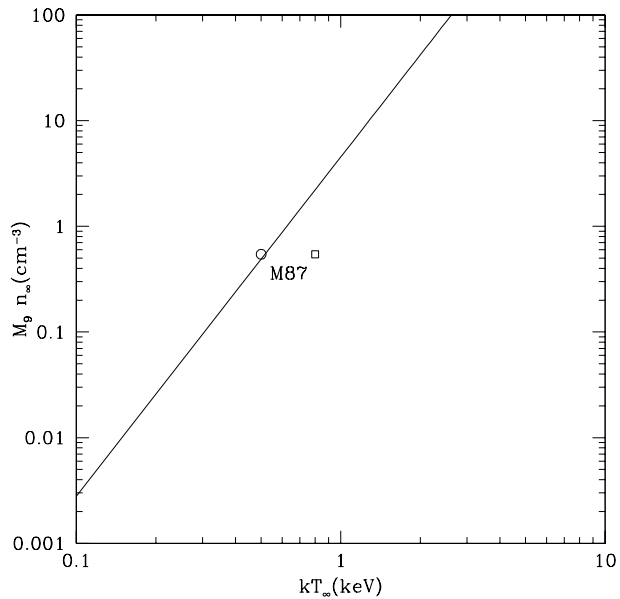,
        angle=0,
        width=3.3in
}
%\noindent
\label{fig:1}
\caption{Parameter space for the balance of heating and cooling 
  at the Bondi radius. The solid line shows the condition when the
  heating rate due to adiabatic compression and the cooling of 3 times
  solar metallicity gas are in balance, so that an accretion solution
  with $\gamma=1$ would be a reasonable approximation. Systems below
  this line are unable to cool to maintain the isothermal condition 
  and therefore $\gamma>1$. The reported data for M87 are shown
  by the square (Di Matteo et al. 2003). A power-law extrapolation of
  the observed temperature gradient with radius to the location of the
  Bondi radius leads to the position marked with a circle. Given
  existing observational data, consideration of a range of equations
  of state, from $\gamma\simeq 1-1.5$ seems appropriate.}
\end{figure}

Now as a specific example, consider M87. It is thought to harbor a
black hole of mass $3.2\pm 0.9\times 10^9\:M_\odot$ (this is the
dynamical mass inside $\sim 5$~pc, Macchetto et al.  1997, but from
the mass-to-light ratio stars are unlikely to contribute significantly,
Harms et al. 1994), which is embedded in gas with a density near the
Bondi radius of about $0.17\pm0.01\:{\rm cm^{-3}}$ and temperature
$0.80\pm0.01\:{\rm keV}$ (Di Matteo et al. 2003). This temperature is
an average of the inner kpc, which is a much larger scale than the
Bondi radius.  Using these values the heating rate is a factor of
14.6, 4.1 times larger than the cooling rate for $\gamma=3/2,1$,
respectively (see also Figure 1).  However the data shown by Di Matteo
et al. (2003) show a temperature gradient so that the inner regions
are cooler. If this trend continues all the way to the Bondi radius
then the cooling rate would balance heating, even keeping the same
value of the density.  Thus the existing data are not able to clearly
distinguish whether conditions are closer to isothermal or adiabatic,
and we must consider a range of possible values of $\gamma$. For the
M87 system, the expected accretion rate is $0.036 f_{\rm B}
(\lambda_\gamma/0.272) M_\odot\:{\rm yr}^{-1}$ and the accretion
luminosity is $2.07 \pm 1.2 \times 10^{44} \epsilon_{0.1} f_{\rm B}
(\lambda_\gamma/0.272)\:{\rm ergs\:s^{-1}}$.

The observed spectrum ($\nu F_\nu$) from the nucleus of M87 peaks
somewhere in the vicinity of $10\:{\rm \mu m}$ (e.g. Di Matteo et
al. 2003), but with large uncertainties. We can estimate the bolometric
luminosity as $L_{\rm bol} \simeq 4\pi d^2 \nu F_\nu \sim 1.4\times
10^{41}\:{\rm ergs\:s^{-1}}$, using the observed $10.8 \:{\rm \mu m}$
flux of $F_\nu = 16.7\pm0.9\:{\rm mJy}$ (Perlman et al.  2001a) and
assuming a distance of 16~Mpc (Tonry et al.  2001). Biretta, Stern, \&
Harris (1991) estimated a total core luminosity from the radio to the
X-ray of $9.1\times 10^{41}\:{\rm ergs\:s^{-1}}$. Note, however, this
estimate is mainly based on an interpolation of flux from the radio to
the optical bands, and is thus quite uncertain. Applying a similar
interpolation for several knots of emission in the jet, Biretta et
al. (1991) estimate it is radiating at $\sim 2\times 10^{42}\:{\rm
ergs\:s^{-1}}$. Note the directly observed luminosities of the core
and jet are only $\sim 10^{41}\:{\rm ergs\:s^{-1}}$ in each of the
optical and X-ray bands (Biretta et al. 1991; Perlman et al. 2001b;
Marshall et al. 2002).

The above luminosity estimates ignore kinetic energy that is escaping
from the inner ($\sim 20\arcsec$) region. Including this contribution,
Reynolds et al. (1996b) estimate that the total power of the nucleus
may be about an order of magnitude greater, i.e. $2\times
10^{43}\:{\rm ergs\:s^{-1}}$, although this is sensitive to the
$10^6\:{\rm yr}$ timescale adopted for the 5~kpc radio halo.

Thus although the total luminosity is quite uncertain, both because of
the limited spectral coverage for the radiated part, and the
difficulty of estimating the kinetic part, it appears that the nucleus
is underluminous by about an order of magnitude with respect to models
of Bondi accretion with $\epsilon=0.1$. 

Note that it is also possible that the nuclear activity is varying on
timescales shorter than the dynamical timescale at the Bondi radius,
$t_{\rm dyn}=r_B/c_{s,\infty}=6.6\times 10^4 \mu_{0.6,\infty}^{3/2}M_9
T_{\rm keV,\infty}^{-3/2}\:{\rm yr}$, which is $\sim3\times 10^5\:{\rm
yr}$ for M87. See Waters \& Zepf (2005) for a discussion on the
flaring of one of the knots of the jet on timescales of just a few
years. M87 is the only one of the ellipticals studied by Di Matteo et
al. (2000, 2003) that is observed to have a powerful jet at the
present time.

%It should be noted however, that unlike the other quiescent
%ellipticals studied by Di Matteo et al. (2000,2003), M87 has a
%powerful jet.  The mechanical luminosity, ultimately supplied by
%accretion, has been estimated to be as large as $2 \ts 10^{43}{\rm
%ergs\:s^{-1}}$ (Reynolds et al. 1996), although with large
%uncertainties.  This would leave only a factor of 10 difference
%between our estimate of the Bondi accretion rate and the accretion
%rate onto the central black hole for the specific case of M87.  

%In
%this respect, M87 is the least underfed of the underluminous
%ellipticals, but it is also the best studied observationally. Thus it
%will be it is very helpful to compare our predictions with this case.

In the next section we propose a new mechanism that helps explain why
the central black holes of massive ellipticals are underluminous: they
are underfed compared to their Bondi rates because much of the
accretion flow condenses into stars. Our mechanism is complementary to
(not contradictory to) any additional low luminosity accretion mode or
outflow mechanism that is operating in the immediate vicininty of the
black hole.

\section{The Accretion Disc}

The accreting material is likely to have an overall net angular
momentum with respect to the central black hole. While the inflow is
subsonic it is relatively easy for turbulent motions, which themselves
are limited to about the sound speed, to transport angular momentum
outwards. This has been observed in the numerical simulations of Abel,
Bryan \& Norman (2002) in the context of primordial star formation
from gas that is only able to cool quite slowly. Following the
approach of Tan \& McKee (2004) we assume that angular momentum is
conserved inside the sonic point $r_{\rm sp}=r_{\rm
  B}(5-3\gamma)/(4\gamma)$ for $\gamma \leq 5/3$, and that the
material passing through this point has a mean circular velocity
$v_{\rm circ} (r_{\rm sp})=f_{\rm Kep} v_{\rm Kep}(r_{\rm sp})$. The
density at the sonic point is $\rho_{\rm
  sp}=\rho_\infty[2/(5-3\gamma)]^{1/(\gamma-1)}$.  The simulations of
Abel et al. (2002) suggest that the circular velocities are
approximately equal to the sound speed, and adopting this as a
fiducial value we have $f_{\rm Kep}=(2\gamma)^{-1/2}$.  If this is a
typical value for the infalling gas, then the median radius at which
material joins a rotationally supported disc is
\beq
r_d= f_{\rm Kep}^2r_{\rm sp} = \frac{r_{\rm sp}}{2\gamma}=r_{\rm B}\frac{5-3\gamma}{8\gamma^2}.
\label{rd}
\ee Depending on the details of the angular momentum distribution,
material will be joining the disc over a range of radii. One possible
description of this assumes uniform rotation at the outer boundary,
which was adopted by Terebey, Shu \& Cassen (1984) and Tan \& McKee
(2004). The estimate implied by eq.~(\ref{rd}) is crude: the actual
median radius depends on the details of the angular momentum
distribution of the infalling gas. The outer radius of the disc may be
several times larger than the median, possibly as large as the Bondi
radius.

\subsection{Gravitational Instability and the Inevitability of Star Formation}

What is the fate of material that becomes centrifugally supported at
$\sim r_d$? For typical conditions, the densities are high enough so
that cooling is sufficient to allow the material to settle into a thin
disc. Local viscous processes and/or global instabilities can then act
to transport angular momentum outwards and matter inwards. Using
the ``alpha'' formalism (Shakura \& Sunyaev 1973) for viscosity $\nu =
\alpha c_{s,d} h = \alpha c_{s,d}^2 / \Omega$, where $h=c_{s,d}/\Omega$ is the
disc half thickness and $\Omega = (GM/r^3)^{1/2}$ is the orbital
angular velocity, the conservation of mass and angular momentum in a
locally steady, thin disc imply the following relation between the
mass accretion rate, viscosity and disc surface mass density, $\Sigma_d
= 2 h \rho_d$,
\beq
\dot{M}_d = 3 \pi \nu \Sigma_d = 3 \pi \alpha \beta_d^b 
c_{s,d}^2 \Omega^{-1} \Sigma,
\label{mdotviscosity}
\ee 
where subscript $d$ refers to conditions in the disc and
$b$ is a parameter that allows viscosity to be proportional to gas
pressure ($b=1$) or total pressure ($b=0$) (Goodman 2003). Note that
use of the relation $h=c_{s,d}/\Omega$ does not account for possible
pressure support from radiation pressure from a young, massive stellar
population in the disc that could couple to the gas via dust.  The
magnetorotational instability can yield values of $\alpha \sim 10^{-3}
- 10^{-1}$ (Balbus \& Hawley 1998), while local self-gravity can give
$\alpha \la 0.3$ (Gammie 2001).  If these processes are relatively
inefficient then for a given accretion rate the disc surface density
builds up to such a high level that the disc becomes gravitationally
unstable and fragments into bound structures that eventually lead to
the formation of stars. This condition can be expressed in terms of
the dimensionless Toomre (1964) parameter \beq Q = \frac{c_{s,d}
  \Omega}{\pi G \Sigma_d} \simeq \frac{\Omega^2}{2\pi G \rho_d},
\label{Q1}
\ee where gravitational instability occurs for $Q<1$.

Combining eqs. (\ref{mdotviscosity}) and (\ref{Q1})
 (Goodman 2003)  we have
\beq
G\dot{M}_d Q = 3 \alpha \beta_d^b 
c_{s,d}^3
\label{Q2}
\ee for a Keplerian disc. The numerical factor 3 becomes $2\sqrt{2}$
for a flat rotation curve.
Although radiation pressure should be small
compared to gas pressure at $r_d$, justifying the assumption that 
$\beta_d\simeq 1$, stellar feedback on dusty gas could provide
partial pressure support analogous to that of radiation pressure.
Here we do not consider this further and set $\beta_d=1$.

First consider a disc that is heated purely by viscous dissipation and
is cooled by radiation from an optically thick surface. The effective
temperature is \beq
\begin{array}{r}
T_{\rm eff,d} = \left(\frac{3}{8\pi \sigma}\frac{GM\dot{M}}{r^3}\right)^{1/4}\\
=4.63 \left[\frac{f_{\rm B}^2 n_\infty^2 T_{\rm keV,\infty}^3}{\mu_{0.6,\infty}} \left(\frac{\lambda_\gamma}{0.272}\right)^2 \left(\frac{8\gamma^2}{5-3\gamma}\right)^6 \right]^{1/8} \left(\frac{r}{r_d}\right)^{-3/4}{\rm K},
\end{array}
\label{teff}
\ee
where $\sigma$ is the Stefan-Boltzmann constant and we have
used Eq. (\ref{mdot}).  
If the disc is optically thick (an assumption that we check later)
then the midplane temperature is given by
\beq
T_d^4 \simeq \frac{3\kappa_d\Sigma_d}{8} T_{\rm eff,d}^4,
\label{Tmid}
\ee assuming the heating rate per unit mass and the opacity per unit
mass are constant. As seen from eq.(\ref{teff}) we are likely to be in
a regime where dust grains can form and survive (i.e. $T_d\la 2000$~K).
The Rosseland mean opacity per unit gas mass then depends on $T_d$ as
well as the gas-to-dust ratio and the properties of the grains. For
example at 100~K and for a gas-to-dust ratio similar to the Milky
Way's the opacity is about $1\:{\rm cm^2\:g^{-1}}$ (Li \& Draine 2001).

The value of the disc surface mass density is
\beq
\begin{array}{r}
\Sigma_d = \frac{\dot{M}_d\Omega}{3\pi \alpha c_{s,d}^2}
 = 153 \alpha_{0.3}^{-4/5} \left(f_{\rm B} n_\infty \frac{\lambda_\gamma}{0.272}\right)^{3/5} \mu_{0.6,\infty}^{9/10} M_9^{4/5}\\ \times T_{\rm keV,\infty}^{-3/20} \kappa_d^{-1/5} \mu_d^{4/5} \left(\frac{8\gamma^2}{5-3\gamma}\right)^{3/5} \left(\frac{r}{r_d}\right)^{-3/5} \:{\rm g\:cm^{-2}},
\end{array}
\label{Sigma}
\ee
where $\mu_d$ is the mean particle mass in the disc in units of $m_{\rm H}$.
The midplane temperature is
\beq
\begin{array}{r}
T_d = 12.7 \left(\frac{\kappa_d \mu_d M_9}{\alpha_{0.3}}\right)^{1/5} \left(f_B n_\infty \frac{\lambda_\gamma}{0.272}\right)^{2/5} \mu_{0.6,\infty}^{1/10} T_{\rm keV,\infty}^{27/80}\\ \times \left(\frac{8\gamma^2}{5-3\gamma}\right)^{9/10} \left(\frac{r}{r_d}\right)^{-9/10}\:{\rm K}.
\end{array}
\label{Tmid2}
\ee Note that this is the minimum temperature that an optically thick
disc can have, since the calculation assumes only viscous heating.
Including the factors that depend on $\gamma$, the fiducial
temperatures are $321,78\:{\rm K}$ for $\gamma=3/2,1$.  Such a disc is
indeed optically thick ($\tau_d=\Sigma_d\kappa_d > 1$), if dust can form that
has properties similar to typical Milky Way dust grains.

What is the value of $Q$ for the optically thick disc? Expressing
equation (\ref{Q2}) in terms of the fiducial parameters of Bondi
accretion in a typical giant elliptical galaxy, we have \beq
\begin{array}{r}
Q=4.86\times 10^{-4} \mu_d^{6/5} \kappa_d^{13/40} \alpha_{0.3}^{7/10} f_{\rm B}^{-2/5} \left(\frac{\lambda_\gamma}{0.272}\right)^{-2/5}
n_\infty^{-2/5}\\ \ts \mu_{0.6,\infty}^{-47/20} M_9^{-17/10} T_{\rm keV,\infty}^{291/160} \left(\frac{8\gamma^2}{5-3\gamma}\right)^{27/20} \left(\frac{r}{r_d}\right)^{-27/20}.
\end{array}
\label{Q3}
\ee The disc is likely to be unstable with respect to its self
gravity: for example applying the parameters appropriate to M87 yields
$Q=(0.600,0.0175) \alpha_{0.3}^{7/10} f_B^{-2/5} \kappa_d^{13/40}$ for
$\gamma=3/2,1$. Note the relatively large value of $Q$ in the
$\gamma=3/2$ case is a consequence of the small value of $r_d$. If a
radius closer to the observed disc size in M87 ($\sim 100$~pc) is
adopted then $Q\ll 1$.

In summary, the accretion of gas to supermassive black holes in the
centers of elliptical galaxies is likely to lead to the formation of a
gravitationally unstable gas disc. The natural expectation is that
this disc will fragment into bound objects: stars. We consider this
process and its consequences in the next section.

\section{Star-Forming Discs Around Galactic Black Holes}

We expect that gravitational instability in the disc leads to
fragmentation and star formation. This process removes mass from the
gaseous disc, reducing $\Sigma_d$. The newly formed stars heat, and
thus stabilize, the remaining gas.

First consider the reduction in the surface density of gas due to star
formation. If the stars remain in a thin disc, the system will still
be prone to gravitational instabilities. These will heat the stellar
population relative to the gas since the latter can cool. As the
scaleheight of the stellar distribution becomes larger than that of
the gas, the stars have less and less influence on the gravitational
stability of the gas. Furthermore, for continued star formation, the
gas itself must be self-gravitating, not just the star-gas system.
This effect can be seen in the star-forming discs of typical disc
galaxies: the threshold for star formation is still adequately
described by the Toomre parameter, $Q$, that depends on the total
surface density of gas (Martin \& Kennicutt 2001). The presence of the
more massive stellar disc raises the particular value of $Q$ below
which star formation occurs, but only by factors of a few or less (Jog
\& Solomon 1984). One way that this can occur is via the local
concentration of gas in stellar spiral arms (e.g. Kuno et al. 1995).
In summary, as $\Sigma_d$ is reduced via star formation, the gas disc
becomes more and more stable with respect to its self-gravity.

An additional effect of the reduction of $\Sigma_d$ is a smaller
optical depth for the disc's cooling radiation. Cooling becomes most
effective when $\tau_d\sim 1$. Other things being equal, we would then
expect a somewhat cooler disc midplane temperature than in the
optically thick limit, given above. However, now the disc is being
heated by stellar photospheres that are much hotter than $T_d$. For a
uniform distribution of gas over the face of the disc, the optical
depth through the disc to this radiation from dust absorption is still
large, even if the optical depth to thermal radiation at $T_d$ is
$\sim 1$.

The amount of heating per mass of new stars depends on their initial
mass function (IMF). The upper part of the IMF appears to
approximately follow a Salpeter (1955) distribution, $d{\cal N}/d m
\propto m^{-2.35}$, for $1\:M_\odot \la m \la 100 M_\odot$  with a
dearth of stars beyond the upper mass limit. The IMFs in the Galaxy
(Salpeter 1955; Kroupa 2002) and in metal-rich starbursts (Schaerer et al. 2000)
appear to be similar in this regime. Below a solar mass the Galactic
IMF flattens somewhat (e.g. Muench et al. 2002), but much of the total
mass is still contained in sub-solar-mass stars. There are no
observational constraints on this part of the IMF in starburst
galaxies.  

For the above IMF, while most of the mass is in low-mass stars, most
of the energy injection into the ISM comes from high-mass stars.
However the value of the high-mass cutoff in the IMF does not
significantly affect the overall feedback properties of the
population, so long as it is greater than $\sim 60M_\odot$ or so.
The uncertainty in the low-mass end introduces an overall
uncertainty in the normalization of feedback per unit stellar mass of
factors of a few: for example if the Salpeter IMF extends down to
$0.1\:M_\odot$ then the feedback per stellar mass is reduced by a
factor of 0.39 from the case with a lower-mass cutoff of $1\:M_\odot$.

Goodman \& Tan (2004) have argued that gravitational instability in
the inner regions of quasar accretion discs (at radii of a fraction of
a parsec) may lead to very massive stars. Conditions there are very
different from those at a typical value of $r_d$ of the Bondi-fed
discs considered in the present paper. In particular, the inner quasar
discs are closer to being dominated by radiation pressure as opposed
to gas pressure; temperatures are too high to allow the presence of
dust grains; and these regions are only marginally self-gravitating
(by definition). We expect that these differences are important enough
to lead to very different IMFs. In short, given existing data we feel
that adopting a Salpeter IMF is the most reasonable approximation to
make for star formation in Bondi-fed gas discs of ellipticals.

For a Salpeter IMF from 0.1 to 100~$M_\odot$, the luminosity
associated with a given star formation rate, $\dot{M}_*$, is $L_*
\simeq 3.9 \times 10^{9} (\dot{M}_*/M_\odot\:{\rm yr^{-1}}) L_\odot =
1.5\times 10^{43}(\dot{M}_*/M_\odot\:{\rm yr^{-1}})\:{\rm
  ergs\:s^{-1}}$ (Leitherer et al. 1999).\footnote{This is actually
  the luminosity of a stellar population that has been continuously
  forming stars at $1.0\:M_\odot\:{\rm yr^{-1}}$ for the last
  $10^7$~yr, which is approximately the age at which ionizing feedback
  reaches a constant level.} After about 30~Myr, this stellar
population produces a supernova at a rate of $7.8\times
10^{-3}(\dot{M}_*/M_\odot\:{\rm yr^{-1}})\rm yr^{-1}$, causing a mean
injection rate of mechanical energy of $\sim 2.3\times
10^{41}(\dot{M}_*/M_\odot\:{\rm yr^{-1}})\:{\rm ergs\:s^{-1}}$. The
mechanical luminosity of stellar winds is estimated to be about a
factor of 6 smaller. Thus the maximum energy input from stars is about
$\epsilon_* \equiv E_*/(M_* c^2) = 2.6\times 10^{-4}$.  The number of
H-ionizing photons produced is $(9.2, 7.9, 7.2)\times
10^{52}(\dot{M}_*/M_\odot\:{\rm yr^{-1}})\:{\rm s^{-1}}$ for
metallicity $Z=1.0, 2.0, 3.0 Z_\odot$ (Smith, Norris, \& Crowther
2002), where the highest metallicity value is based on a power law
extrapolation from the lower two values.  Assuming case B
recombination in the ionized regions with no escape of ionizing
photons yields an H$\alpha$ luminosity of (Kennicutt, Tamblyn, \&
Congdon 1994) \beq L_{\rm H\alpha} = 0.99 \times 10^{41} f_{\rm
  trap,i} (\dot{M}_*/M_\odot\:{\rm yr^{-1}}) \:{\rm ergs\:s^{-1}},
\label{sfr}
\ee with the normalization being for the $Z=3Z_\odot$ case and where
$f_{\rm trap,i}$ is the fraction of ionizing photons that are trapped
in the disc's HII region and do not escape into the hot interstellar
medium of the galaxy. This estimate ignores dust absorption in the
disc, which would lower the numerical factor in the above equation.
Note that a recombination flow of one solar mass per year would
produce an H$\alpha$ luminosity that is a factor of $4.4\times
10^{-4}$ times smaller than the amount produced by one solar mass per
year of star formation (with the above Salpeter IMF).

The H$\alpha$ flux from the circumnuclear disc of M87 is
$4.4\pm1.5\ts10^{-14}\:{\rm ergs\:cm^{-2}\:s^{-1}}$ (Ford et al.
1994) (see Figure 2), corresponding to a luminosity of $1.35\ts
10^{39} \:{\rm ergs\:s^{-1}}$ for an assumed distance of 16~Mpc (Tonry
et al.  2001). This implies a star formation rate of $1.4\ts 10^{-2}
f_{\rm trap,i}^{-1}\:M_\odot \:{\rm yr^{-1}}$ using the normalization
of equation~(\ref{sfr}). Note that this estimate is a lower
limit because it does not allow for any internal extinction in the
star-forming disc that would attenuate the observed H$\alpha$ flux.

If we assume that the bolometric luminosity of the circumnuclear disc
in M87 is similar to the observed peak in $\nu L_\nu$, i.e. $\sim
1.4\ts 10^{41}\:{\rm ergs\:s^{-1}}$ (Perlman et al. 2001a; see \S2),
and is due entirely to star formation, then the above relations imply
a star formation rate of $9.5\times 10^{-3}\:M_\odot \:{\rm yr^{-1}}$.
Figure 3, described in more detail in the next section, shows how the
bolometric luminosity due to star formation compares to the observed
spectrum of the M87 nucleus.

Despite the uncertainties in the IMF, internal extinction and equation
of state of the accreting gas, the key result from the above star
formation rate estimates is that they are both roughly consistent with
the inferred Bondi accretion rate. This agreement suggests that the
star-forming accretion flow model provides a plausible mechanism for
preventing significant gas accretion onto the central black hole, at
least in M87.
%We regard the agreement of these
%estimates of the star formation rate with the Bondi feeding rate of
%the disc to be strong pieces of evidence in support of the model, at
%least in the case of M87. 
This is in addition to the fact that a relatively thin
disc is observed to be present on scales just inside the Bondi radius
(see further discussion in \S5.1).

\begin{figure}
\label{fig:disc}
\epsfig{file=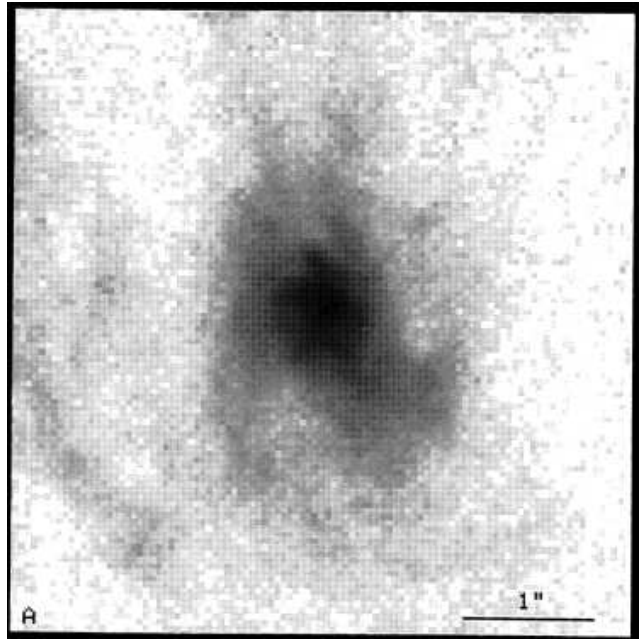,
        angle=0,
        width=3.3in
}
%\noindent
\caption{Adapted from Ford et al. (1994): HST F658N on-band image minus HST F547M off-band image of the M87 nucleus designed to highlight H$\alpha$ emission. North is towards the top and east to the left; the famous synchrotron jet (not visible here) extends to the NW. The black hole is at the center of the image and is surrounded by a disc of material: the velocity difference from 1\arcsec\, to the NE to 1\arcsec\, to the SW is 1000~$\rm km\:s^{-1}$ (Harms et al. 1994; Macchetto et al. 1997). The presence of this disc, including its spiral structure and dust lanes, as well as its total H$\alpha$ luminosity, is evidence in support of a model of circumnuclear star formation.
}
\end{figure}

\subsection{Estimate of the Star Formation Efficiency and Residual Accretion to the Black Hole}

Here we present a simple one zone model that describes the star
formation efficiency of a $Q=1$ disc as a function of the coupling of
star formation feedback to the gas. From equation~(\ref{Q2}) we have
\beq
\label{Qint}
Q = \frac{3\alpha}{G\dot{M}_d} \left(\frac{kT_d}{\mu_d}\right)^{3/2}.
\ee For an optically thick disc, $T_d^4\simeq 3 \kappa_d \Sigma T_{\rm
  eff}^4 / 8$ and its luminosity is $L \simeq 2 \pi r_d^2 \sigma
T_{\rm eff}^4 = f_{\rm trap,L} L_* \simeq f_{\rm trap,L}
3.9\ts10^{9} (\dot{M}_*/M_\odot\:{\rm yr^{-1}}) L_\odot$, assuming a
negligible contribution from viscous heating, valid at $r\sim r_d$ if
we are to achieve a quasi-stable disc with $Q\sim 1$.  In the last
step above, we have used the IMF-dependent relation between luminosity
and star formation rate discussed previously. Now by using the above
relations, including eq.~(\ref{Q1}), assuming that $Q=1$ and that most
of the Bondi accretion rate goes into stars, i.e.  $\dot{M}_*\simeq
\dot{M}$, we derive the mass accretion rate in the disc that can be
supported against self-gravity by stellar feedback: \beq
\label{mdotdisc}
\begin{array}{r}
\dot{M}_d = 2.27 \times 10^{-6} \alpha_{0.3} \left(\frac{\kappa_d^2 f_{\rm trap,L}^2 L_1^2 M_9}{\mu_d^8}\right)^{3/14} \\ \ts \left(\frac{\dot{M}_*}{M_\odot\:{\rm yr^{-1}}}\right)^{3/7} \left(\frac{r_d}{100\:{\rm pc}}\right)^{-3/2}\:M_\odot\:{\rm yr^{-1}}\\
= 2.686 \times 10^{-6} \alpha_{0.3} \left(\frac{\kappa_d f_{\rm trap,L} L_1 T_{\rm keV,\infty}^2 f_{\rm B} n_{\infty}}{\mu_d^4 M_9 \mu_{0.6,\infty}}\frac{\lambda_\gamma}{0.272}\right)^{3/7}\\
\ts  \left(\frac{8\gamma^2}{5-3\gamma}\right)^{3/2}\:M_\odot\:{\rm yr^{-1}},
\end{array}
\ee where $L_1=L_*/3.9\times 10^{9}L_\odot$ is the luminosity
generated by a star formation rate of $1\:M_\odot\:{\rm yr^{-1}}$ for
our adopted IMF.  This is the residual amount of gas accretion that
can be stabilized with respect to its own self-gravity by transforming
most of the initial mass flux into stars. We have written
eq.~(\ref{mdotdisc}) in two stages so that different constraints can
be used: either the observed disc size and the Bondi accretion rate to
the disc (or any general mass feeding rate to the disc), or the
properties of the hot gas in the elliptical (in this case the
theoretical estimate for the disc size is used; eq.~\ref{rd}).  For M87, in the
first case the accretion rate in the disc is $(7.0\times 10^{-7}, 1.3
\times 10^{-6}) \alpha_{0.3} (\kappa_d f_{\rm trap,L} f_B
\mu_d^{-4})^{3/7} M_\odot\:{\rm yr^{-1}}$ for $\gamma=3/2,1$ with
$r_d=100$~pc, while in the second case it is $(1.36\times 10^{-4},
9.25 \times 10^{-6}) \alpha_{0.3} (\kappa_d f_{\rm trap,L} f_B
\mu_d^{-4})^{3/7} M_\odot\:{\rm yr^{-1}}$ for $\gamma=3/2,1$. These
are much lower than the initial Bondi accretion rates of $(0.036,
0.148) f_{\rm B}\:M_\odot\:{\rm yr^{-1}}$.

If the disc is optically thin to its cooling radiation, then
$T_d^4\simeq T_{\rm eff,d}^4 / \tau_d = T_{\rm eff,d}^4 /(\kappa_d \Sigma_d)$,
and the equation for the mass accretion rate in the disc becomes
\beq
\label{mdotdiscthin}
\begin{array}{r}
\dot{M}_d = 8.57 \times 10^{-5} \alpha_{0.3} \left(\frac{f_{\rm trap,L}^2 L_1^2}{\mu_d^8 \kappa_d^2 M_9}\right)^{1/6} \\ \ts \left(\frac{\dot{M}_*}{M_\odot\:{\rm yr^{-1}}}\right)^{1/3} \left(\frac{r_d}{100\:{\rm pc}}\right)^{-1/6}\:M_\odot\:{\rm yr^{-1}}\\
= 2.63 \times 10^{-5} \alpha_{0.3} \left(\frac{f_{\rm trap,L} L_1 f_{\rm B} \mu_{0.6,\infty}^2 n_{\infty} M_9}{\mu_d^4 \kappa_d T_{\rm keV,\infty}} \frac{\lambda_\gamma}{0.272}\right)^{1/3}\\
\ts  \left(\frac{8\gamma^2}{5-3\gamma}\right)^{1/6}\:M_\odot\:{\rm yr^{-1}}.
\end{array}
\ee Now for M87, in the first formula of (\ref{mdotdiscthin}) the
accretion rate in the disc is $(2.3\times 10^{-5}, 3.7 \times 10^{-5})
\alpha_{0.3} (\kappa_d^{-1}f_{\rm trap,L} f_B \mu_d^{-4})^{1/3}
M_\odot\:{\rm yr^{-1}}$ for $\gamma=3/2,1$ with $r_d=100$~pc, while in
the second formula it is $(4.2\times 10^{-5}, 4.7 \times 10^{-5})
\alpha_{0.3} (\kappa_d f_{\rm trap,L} f_B \mu_d^{-4})^{3/7}
M_\odot\:{\rm yr^{-1}}$ for $\gamma=3/2,1$.

We can judge which of the optically thin or thick limits is
appropriate by the value of $\Sigma_d = \dot{M}_d \Omega /(3\pi \alpha
c_{s,d}^2)=(\dot{M}_d/(3\alpha))^{1/3}\Omega/ (\pi (GQ)^{2/3})$ and the
disc temperature, which sets $\kappa_d$. For the optically thick case
the surface density is \beq
\label{Sigmathick}
\begin{array}{r}
\Sigma_d = 7.05\times 10^{-3} \left(\frac{\kappa_d f_{\rm trap,L} L_1 M_9^4}{\mu_d^4}\frac{\dot{M}_*}{M_\odot\:{\rm yr^{-1}}}\right)^{1/7}\\ \ts \left(\frac{r_d}{100\:{\rm pc}}\right)^{-2}\:{\rm g\:cm^{-2}},
\end{array}
\ee
while in the optically thin case it is
\beq
\label{Sigmathin}
\begin{array}{r}
\Sigma_d = 0.0236 \left(\frac{f_{\rm trap,L} L_1 M_9^4}{\kappa_d \mu_d^4}\frac{\dot{M}_*}{M_\odot\:{\rm yr^{-1}}}\right)^{1/9}\\ \ts \left(\frac{r_d}{100\:{\rm pc}}\right)^{-14/9}\:{\rm g\:cm^{-2}}.
\end{array}
\ee For a dust-to-gas ratio three times that of the Milky Way's,
$A_V\simeq 600 \Sigma$ so that the above two numerical
factors correspond to $A_V=4.2,14$, respectively. The effective
temperature at the disc surface is \beq
\label{Teff}
\begin{array}{r}
T_{\rm eff,d} = \left(\frac{f_{\rm trap,L} L_*}{2\pi r_d^2 \sigma}\right)^{1/4} = 25.8 \left(\frac{f_{\rm trap,L}L_1\dot{M}_*}{M_\odot\:{\rm yr^{-1}}}\right)^{1/4}\\ \ts \left( \frac{r_d}{100\:{\rm pc}}\right)^{1/2}\:{\rm K}
\end{array}
\ee For most elliptical galaxies, the disc will be optically thin to
its cooling radiation in the outer parts, and will make a transition
to the optically thick regime in the inner region. Note that the discs
are expected to be quite optically thick to their stellar radiation
field, although clustering of young stars and ISM inhomogeneities will
act to reduce this. Because of these complications we have left the
factor $f_{\rm trap,L}$ as a free parameter, but expect it to have a
value of order unity.

Despite the uncertainties in the parameters and the dependence on the
adopted disc size, the estimates suggest that only a small fraction of
the initial mass accretion flux through the disc would be supported
against star formation and accreted by the central black hole. The
amount of the reduction implied by equations~(\ref{mdotdisc}) and
(\ref{mdotdiscthin}) is more than sufficient to explain the low
luminosity of M87's nucleus. However, the residual accretion rates
predicted by the above simple model are so small that other processes
that have not been considered, such as mass supply from stellar winds,
may now be relatively important. Another caveat is that the above
estimates only consider thermal pressure support in the gas disc and
ignore the additional stabilizing effect of magnetic fields.

Note that some material will be expelled from stars as they undergo
and complete their stellar evolution. For Salpeter-like IMFs, most of
this matter will be returned after a relatively long timescale: for
our adopted Salpeter IMF from 0.1 to 100~$M_\odot$, only 14\% of the
mass is in stars with masses greater than $8\:M_\odot$. Irrespective
of the timescale of mass return, most of this material would be
recycled back into the gas disc rather than accreted directly by the
black hole. This is because most stars are forming and evolving at
distances from the black hole that are still quite large compared to
the sizes of their stellar wind bubbles or planetary nebulae, and the
specific angular momentum of the returned gas is the same as that of
the stars.

Other authors have highlighted the issue of self-gravity in the
accretion discs that feed supermassive black holes, mainly in the
context of quasars that are accreting at much higher rates,
approaching their Eddington limits (Schlosman \& Begelman 1989;
Goodman 2003; Sirko \& Goodman 2003; Thompson et al. 2005). Goodman
(2003) pointed out that stellar heating of the disc was unlikely to be
able to stabilize the disc, because the energy release per mass of
stars formed is relatively low.  We have extended this line of
reasoning to conclude that most of the mass flux goes into stars,
which are then able to stabilize a small residual fraction of the gas
in an accretion disc to the black hole. The implications of this model
for the fueling of quasars are discussed in \S5.4.

\subsection{Estimate of the total mass of gas in the disc}

One prediction of our model of (Bondi-fed) star-forming accretion
flows is the presence of cold, presumably molecular gas.  To estimate
the amount, we assume that a steady state has been achieved so that
the star formation rate in the disc balances the Bondi accretion rate
to the disc. The disc mass is $M_d \simeq \pi r_d^2 \Sigma_d$. In the
optically thick case this is \beq
\label{mdthick}
M_d = 1.06\times 10^6 \left(\frac{\kappa_d f_{\rm trap,L} L_1 M_9^4}{\mu_d^4}\frac{\dot{M}_*}{M_\odot\:{\rm yr^{-1}}}\right)^{1/7}\:M_\odot,
\ee
while in the optically thin case it is
\beq
\label{mdthin}
\begin{array}{r}
M_d = 3.55\times 10^6 \left(\frac{f_{\rm trap,L} L_1 M_9^4}{\kappa_d \mu_d^4}\frac{\dot{M}_*}{M_\odot\:{\rm yr^{-1}}}\right)^{1/9}\\ \ts\left(\frac{r_d}{100\:{\rm pc}}\right)^{4/9}\:M_\odot.
\end{array}
\ee For M87 the disc mass is then $(1.3,1.6)\ts 10^6 (\kappa_d f_{\rm
  trap,L} L_1 \mu_d^{-4})^{1/7}\:M_\odot$ ($\gamma=3/2,1$) in the
optically thick case and $(4.1,4.8)\ts 10^6 (\kappa_d^{-1} f_{\rm
  trap,L} L_1 \mu_d^{-4})^{1/9} \:M_\odot$ ($\gamma=3/2,1$) in the
optically thin case.

As an alternative method,
we use an empirical relation between star formation rate
and disc gas mass to estimate the properties of the gas disc.
For a sample of about a hundred galactic and circumnuclear discs,
Kennicutt (1998) found an empirical linear relation between the
surface density of star formation and the product of the total mean
gas surface density and the mean orbital angular velocity: \beq
\label{kennicutt}
\Sigma_{\rm SFR} \simeq 0.017 \Sigma_d \Omega, \ee where $\Sigma_{\rm
  SFR}\equiv \dot{M}_*/(\pi r_d^2)$. It is not clear if this relation
applies to Bondi-fed circumnuclear discs in ellipticals as the orbital
timescales are much shorter than in the sample of Kennicutt and the
rotation curves may be closer to Keplerian (e.g. M87 -- Macchetto et
al. 1997), rather than being flat or rising. This may lead to somewhat
different star formation efficiencies, at least in the context of
shear-dependent models of star formation (Tan 2000).

Nevertheless, using eq.~(\ref{kennicutt}) we can estimate the surface
density of the gas disc: 
\beq
\label{sigmagaskennicutt}
\Sigma_d = 0.184 \frac{\dot{M}_*}{M_\odot\:{\rm yr^{-1}}} M_9^{-1/2} \left(\frac{r_d}{\rm 100\:pc}\right)^{-1/2}\:{\rm g\:cm^{-2}}.
%\Sigma = 5.32\ts 10^{-3} \frac{\lambda_\gamma}{0.272} \frac{n_\infty \mu_{0.6,\infty}^2 f_{\rm B}M_9}{T_{\rm keV,\infty}} \left(\frac{8\gamma^2}{5-3\gamma}\right)^{1/2}\:{\rm g\:cm^{-2}}.  
\ee In the case of M87 this takes values $(3.7,15.3) \ts 10^{-3}\:{\rm
  g\:cm^{-2}} \equiv (18, 73)\:M_\odot\:{\rm pc^{-2}}$ for
$\gamma=3/2,1$, which is in the middle of the range of values covered
by Kennicutt's sample.  The total gas mass in this disc is \beq
\label{mtotkennicutt}
\begin{array}{r}
M_d = \pi r_d^2 \Sigma_d = \frac{\dot{M}_*}{0.017\Omega}\\ = 2.77\ts 10^7 \frac{\dot{M}_*}{M_\odot\:{\rm yr^{-1}}} M_9^{-1/2} \left(\frac{r_d}{\rm 100\:pc}\right)^{3/2}\:M_\odot,
%M_d \equiv \pi r_d^2 \Sigma_{\rm gas} \\= 5.8\ts 10^4 \frac{\lambda_\gamma}{0.272} \frac{n_\infty \mu_{0.6,\infty}^4 f_{\rm B} M_9^3}{T_{\rm  keV,\infty}^3 } \left(\frac{8\gamma^2}{5-3\gamma}\right)^{-3/2} M_\odot, 
\end{array}
\ee which takes values $(0.6, 2.3\ts10^6)\:M_\odot$ in the case of M87
with $\gamma=3/2,1$. The coincidence of the two different methods of
estimating the disc mass, suggests that this feedback model may have
relevance to setting the normalization of the empirical Kennicutt
(1998) relation.

Note that if the disc were disrupted by a particular event, such as an
outburst from the AGN, then it would take $\sim 10^7 - 10^8$~yr to
reform, if supplied at the Bondi rate.

%% We note here that the total mass estimate is
%% sensitive to the actual size of the disc (via $\Omega$), which depends
%% on the initial angular momentum distribution, i.e. $f_{\rm Kep}$. The
%% observed size of the H$\alpha$ disc in M87 is about 100~pc, which is
%% several times larger than the value of $r_d$ predicted by equation
%% (\ref{rd}). 
%% %Using the observed disc size, increases the mass estimates
%% %to $(0.6,2.6)\times 10^6\:M_\odot$, respectively.

%% The efficiency of the coupling of feedback processes to the gas may be
%% relatively small, particularly if star formation is clustered and if
%% the stars can migrate from the gas during their main sequence
%% lifetimes. In normal Galactic star-forming regions, about ?? of O and
%% ?? of B stars are runaways, whose supernovae would only weakly couple
%% to a thin gas disk. 

%% Compared to typical Galactic star forming regions,
%% the temperatures are somewhat elevated, but the pressures are likely
%% to be greater: these effects counter one another in setting the
%% minimum fragment mass, i.e. the thermal Jeans mass at a typical point
%% of core formation. 

%$t_{ff}$ compared to $t_{orbit}$ for Q=1 disk.

%% If the disc is optically thin then $T_d^4 \sim T_{\rm eff}^4/(\kappa
%% \Sigma)$ and the emissivity of the disc surface would be reduced below
%% the black body value.

%Value of Q in optically thin case.

\section{Discussion}

We have presented a model of star-forming accretion flows with
particular attention to the flow inside the Bondi radius of the
central black hole of an elliptical galaxy.  Our approach was as
follows: first we calculated the accretion rate, showing the explicit
dependence on the effective equation of state of the gas. We then
estimated the heating and cooling rates of the gas in the accretion
flow near the Bondi radius, showing that idealized accretion solutions
$\gamma\simeq 1 - 1.5$ are probably quite a reasonable description of
the actual accretion flow, at least in the case of M87.  We then
argued that angular momentum conservation inside the sonic point of
the flow leads to formation of a disc on scales that are a factor of a
few or so smaller than the Bondi radius. The disc was shown to be
gravitationally unstable if the heating is due to viscous processes in
the disc. We found that self-gravity is so strong that most of the
mass flux will be converted into stars. The energy input from these
star can stabilize only a small residual fraction of the initial mass
flux: i.e. the accretion rate to the black hole is substantially
reduced below the Bondi accretion rate. We argue that this is a major
cause of the ``low luminosity'' of black holes observed in giant
elliptical galaxies, where ``low luminosity'' is in reference to a
model of thin disc accretion at the Bondi rate as determined from the
conditions in the kpc-scale, X-ray-emitting gas. A
radiatively-inefficient accretion flow may exist at the center of the
disc, much closer to the black hole. However, its inefficiency must be
argued in the context of an accretion rate and a luminosity determined
at a much smaller scale, for example where the disc is no longer
self-gravitating.

The predictions of this model include the presence of a star-forming
gas disc (i.e. containing relatively cold molecular gas) around the
black holes of ellipticals and the presence of hot, young stars in
these regions.  These may reveal themselves by spectral features (e.g.
clustered H$\alpha$ emission from gas ionized by massive stars;
stellar atmosphere/wind features, e.g. from Wolf-Rayet stars; or
stellar ultraviolet continuum, although this may be swamped by AGN
emission) or by the occurrence of type II supernovae. For example a
typical star formation rate at the center of a giant elliptical galaxy
is expected to be $\sim 0.1\:M_\odot\:{\rm yr^{-1}}$, about a factor
of 30 less than the star formation rate of a Milky-Way-like disc
galaxy. For the standard Salpeter IMF adopted here, the type II
supernova rate from such an elliptical galaxy is $\sim 8\times
10^{-4}\:{\rm yr^{-1}}$.

Nayakshin (2004) considered how the presence of many young stars, left
over from a previous quasar phase, can affect the accretion and
appearance of low-luminosity AGN. However, he did not consider the
possibility that the stars are currently forming in these systems.

We now summarize the potential application of the star-forming
accretion disc model to M87, other giant ellipticals, the Galactic
Center, quasars and other AGN.

\subsection{Application to M87}

We have used M87 as an example to illustrate the model throughout the
text. Here we summarize this application, discuss the existing
evidence that supports the model, and suggest future observations that
can confirm or refute its validity.

The central black hole of M87, with mass $(3.2\pm
0.9)\ts10^9\:M_\odot$ (Macchetto et al. 1997), should have a Bondi
accretion rate of $0.036-0.15\:M_\odot\:{\rm yr^{-1}}$ (this range
corresponds to uncertainties in $\gamma=3/2 - 1$) from the hot
(0.8~keV), low-density ($n_e\simeq0.17\:{\rm cm^{-3}}$) gas (Di Matteo
et al.  2003) at the center of the giant elliptical (uncertainties in
the black hole mass and gas temperature could also change the
accretion rate by factors of three or so). The Bondi radius is about
$110$~pc, but could be twice as large if the high end of the black
hole mass determination is valid and the temperature in this inner
region is actually 0.5~keV (see discussion in \S2). The expected {\it
  median} disc radius is several times smaller than the Bondi radius:
our fiducial case has it smaller by factors of 36 and 4 for
$\gamma=3/2,1$. The actual outer disc radius would be larger than the
median by factors of a few, and in any case the exact size depends on
the details of the initial angular momentum distribution of the hot
accreting gas, which is not well constrained.  Since the H$\alpha$
disc seen by HST is $\simeq 100$pc in extent (Ford et al.  1994; see
Fig.~2), this is probably the best estimate of the disc scale in this
system. Note also that the HST observations show some filamentary
structure out to several hundred pc, so that the accretion process may
not be quite a simple as given by the purely Bondi solution.

Despite the caveats, the coincidence of the observed disc size with
the estimated Bondi radius supports a model in which that disc is
forming from a Bondi accretion flow with initial circular motions of
order the hot gas sound speed.

The H$\alpha $ disc was described by Ford et al. (1994) as having
``spiral structure'' and ``dust lanes'' (see Fig.~2), which supports
the theoretical conclusion that it is gravitationally unstable (\S3)
and contains cool gas that is liable to undergo star formation
(\S\S3,4).  The total H$\alpha$ luminosity implies a star formation
rate (assuming a standard Salpeter IMF) within factors of a few of the
Bondi accretion rate at which the disc is being fed. This is
consistent with the implied star formation rate being equal to the Bondi
accretion rate given the uncertainties.

We can model the bolometric luminosity associated with this star
formation rate.  Figure~3 shows the observed infrared and radio
spectrum of the M87 nucleus, much of which results from nonthermal
emission from the AGN (see \S2). Care must be taken since the fluxes
at different wavelengths often correspond to regions of different
scales. Also shown in Figure~3 is a simple black body spectrum of the
gas and dust disc heated by star formation at rates
$0.036-0.15\:M_\odot\:{\rm yr^{-1}}$, corresponding to luminosities
$(1.6-5.9)\ts 10^8\:L_\odot$. The temperature is set by assuming this
luminosity is emitted over the two faces of a disc of radius
100~pc. This model is highly idealized, since it assumes a single
temperature, ignores additional heating from the interstellar
radiation field of the galaxy, ignores deviations from a pure black
body spectrum, such as PAH emission features in the range $1-20\:{\rm
\mu m}$, and assumes all the light from stars is reprocessed by
dust. The overall luminosity normalization is also
IMF-dependent. Nevertheless it is interesting that these simple models
are now in a position to be tested by high resolution infrared and
sub-mm observations that are now feasible with the Spitzer Space
Telescope and the Sub-millimeter Array (SMA).

\begin{figure}
\label{fig:spec}
\epsfig{file=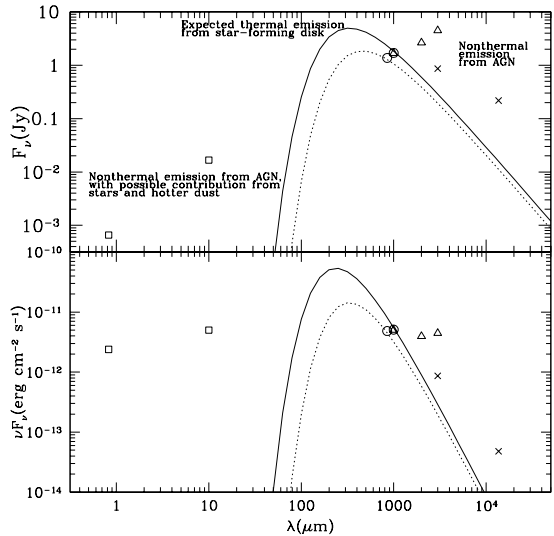,
        angle=0,
        width=3in
}
\caption{Continuum spectrum of the nucleus of M87 shown as $F_\nu$ (top) 
  and $\nu F_\nu$ (bottom). Data in order of increasing wavelength:
  HST optical (Ho 1999); Gemini 10~$\rm \mu m$ (Perlman et al. 2001a);
  Circles --- SMA calibration data (sensitive to flux on scales of the
  order $2\arcsec$); Triangles --- IRAM 30m data (sensitive to flux on
  scales of order $11\arcsec$ and $28\arcsec$ at 1.3\,mm and 3\,mm,
  respectively (data also extracted from the SMA calibrator list);
  Crosses --- VLBI data at 100~GHz (Baath et al. 1992) and 22~GHz
  (Spencer \& Junor 1986) sensitive to flux on scales of $0.2\arcsec$
  and $\sim$0.00015\arcsec. The solid curve is a black body spectrum
  from a star-forming disc with the star formation rate equal to the
  upper limit of the Bondi feeding rate of the disc
  ($0.15\:M_\odot\:{\rm yr^{-1}}$) and a standard Salpeter mass
  function down to $0.1\:M_\odot$. The temperature used here is 16~K
  set by having the emitting area be equal to that of the observed
  H$\alpha$ disc with radius $r_d \simeq 100$~pc. This is a lower
  limit as it neglects additional heating from the evolved stellar
  population of the galaxy. The dotted curve is the equivalent model
  for the lower limit of the Bondi accretion rate
  ($0.036\:M_\odot\:{\rm yr^{-1}}$).
%SMA observations that resolve the
%  disc can help separate thermal disc emission from nonthermal radio
%  emission from the parsec scale region around the black hole.
}
\end{figure}

The presence of cold, presumably molecular gas in the disc with total
mass $\sim 10^6\:M_\odot$ could be searched for with mm
interferometers. Such a search is particularly well suited to the
capabilities of the SMA and ALMA.

%Optical spectra may be able to reveal the presence of hot, young
%massive stars.

To estimate the efficiency of the central engine we need to know the
black hole's accretion rate. As noted in \S2, to explain the
observations requires a reduction in the accretion rate by a factor of
$\sim 10$ from the Bondi value (this factor is itself uncertain by
about an order of magnitude), or a similar reduction in the efficiency
of energy liberation from the fiducial value of $\epsilon=0.1$. Note
that most of the energy is thought to be released as non-radiating
kinetic energy of the jet/outflow.  The reduction in black hole
accretion rate from the Bondi rate due to star formation is also quite
uncertain. We estimated (eq. \ref{mdotdiscthin}) a possible reduction
by factors of up to $\sim 1000$. However, as discussed in \S4, this
simple estimate is based on a simple one-zone model for the disc and
does not include effects such as accretion of some very low angular
momentum gas from the larger-scale galaxy and the return of material
evolved stars. In summary, the observations of the M87 nucleus can be
explained by models of efficient ($\epsilon\sim0.1$) accretion if a
large fraction ($\sim 90\%$) of the initial mass flux to the disc
forms stars. While such a high fraction is certainly allowed by the
simple estimates of self-regulated star formation we have presented in
\S4, quantitative estimates of the residual accretion rate are difficult.

\subsection{Other Elliptical Galaxies}

Optical imaging with HST has been used to detect dusty discs in
elliptical galaxies.  Jaffe et al. (1993) reported a disc of cool gas
and dust with radial extent of $65$~pc around the nucleus of NGC~4261
(3C~270), another elliptical galaxy in the Virgo cluster (see also
Martel et al. 2000 and van Bemmel et al. 2004).  Bower et al. (1997)
reported a disc, with radius 82~pc, of ionized gas in M84 (NGC~4374 =
3C~272.1). Jaffe et al.  (1994) surveyed 14 Virgo ellipticals, finding
clear evidence for dust discs (seen in absorption) in several. Note
this method of detection could be sensitive to the orientation of the
disc and other properties of the nucleus: for example in this study no
dust disc was seen in absorption in the case of M87. They note that
essentially all their systems exhibit a change in morphology on scales
of $\sim 100-200$~pc from the nucleus: either in the extent of an
absorbing dust disc or a change in the brightness profile. We identify
this scale with the Bondi radius, inside which a cool, star-forming
disc can form.
%If the gas
%temperature near the Bondi radius is proportional to the virial
%temperature, then $T_\infty \propto T_{\rm vir} \propto \sigma_*^2
%\propto M_{\rm gal}^{0.32}$, where $\sigma_*$ is the galaxy's stellar
%velocity dispersion and $M_{\rm gal}$ is its mass, and the last step
%uses the empirical relation of Faber et al. (1997). Assuming a
%constant black hole to galaxy mass fraction, i.e. $M\propto M_{\rm
%  gal}$, the Bondi radius then scales as $r_{\rm B}\propto M_{\rm
%  gal}^{0.68}$.
Van Dokkum \& Franx (1995) found that dust ``lanes'' were present in
the majority ($\sim 80\%$) of a sample of 64 early-type galaxies.
These were usually perpendicular to the axis of any radio jet present
in the system.  Carollo et al. (1997) and Tomita et al. (2000) also
found evidence for nuclear dust absorption in the majority of the
galaxies they analyzed. Martel et al. (2004) found nuclear discs of
dust and ionized gas in a large fraction of the nine early type
galaxies they studied with HST-ACS. Again the typical scales of these
features are $\sim 100$~pc. They estimated star formation rates of
$(0.6-6)\times 10^{-3}\:M_\odot\:{\rm yr^{-1}}$ assuming all the
H$\alpha$ emission is produced by star formation and using the
normalization of Kennicutt (1998). Lauer et al. (2005) found dust in
the centers of about half their sample of 77 galaxies. Only one of
their ``dusty'' galaxies had dust that was more prominent on scales
greater than 4\arcsec from the center than closer to the center; most
of the dust is highly concentrated, and in all the systems where the
dust is clearly in a disc or a ring, no dust clouds are visible
external to the outer edge of the disc or ring. Significantly, no dust
was seen in relatively low luminosity, low mass ellipticals
($M_V>-21$), which might be expected since the Bondi accretion rate
increases with galaxy mass, assuming the black hole mass is a fixed
fraction of the galaxy mass.

Jaffe \& McNamara (1994) detected radio absorption in HI and CO from
neutral gas in the disc of NGC~4261 and estimated a mass $\sim 2\times
10^5\:M_\odot$, within an order of magnitude of the theoretical mass
estimates we have made for the disc in M87.  Knapp \& Rupen (1996)
surveyed 42 elliptical galaxies for CO emission, concluding that the
detection rate was just under 50\%. However, since the half-power beam
of these observations was about 30\arcsec, these detections most likely do
not relate to molecular gas forming inside the Bondi radius, but to
material formed from evolved stars or associated with the infall of
gas-rich satellite galaxies. Young (2002) detected CO emission in 5
out of 7 ellipticals, but again, because of the resolution of the
observations, the focus was on $\sim$kpc scales.  Vila-Vilar\'o, Cepa,
\& Butner (2003) detected CO(3-2) emission from 6 out of 10 early-type
galaxies that already showed CO(2-1) emission, with a beam of about
20\arcsec. They inferred typical star formation rates (from H$\alpha$) of a
few tenths of a solar mass per year, but again this is mostly on
scales larger than the Bondi radius (nuclear H$\alpha$ emission was
not considered in this study).
%range $\sim 10^{-3}-10^{-1}\:M_\odot\:{\rm yr^{-1}}$, but, as they describe, 

In summary, there is much evidence that elliptical galaxies contain
dusty gas discs on scales approximately equal to their Bondi radii,
i.e. $\sim 100$~pc. It is less clear if star formation is actively
occurring in these discs. One could make a more quantitative test of
whether central dust concentrations are related to the Bondi radius by
comparing the size of the dust features with the $r_{\rm B}$.
If the gas
temperature near the Bondi radius is proportional to the virial
temperature, then $T_\infty \propto T_{\rm vir} \propto \sigma_*^2
\propto M_{\rm gal}^{0.32}$, where $\sigma_*$ is the galaxy's stellar
velocity dispersion and $M_{\rm gal}$ is its mass, and the last step
uses the empirical relation of Faber et al. (1997). Assuming a
constant black hole to galaxy mass fraction, i.e. $M\propto M_{\rm
  gal}$, the Bondi radius then scales as $r_{\rm B}\propto M_{\rm
  gal}^{0.68}$.

Long term star formation at the scale of the Bondi radius ($\sim
100\:{\rm pc}$) will have implications for the morphologies of the
cores of elliptical galaxies. For example it may account for the $\sim
100$~pc scale discs seen in some early type galaxies (e.g. van den
Bosch, Jaffe, \& van der Marel 1998; Krajnovi\'c \& Jaffe 2004).  It
may also account for the flattening in the radial profiles of ``core''
elliptical galaxies (e.g., Lauer et al. 2005): this occurs on scales
of $\sim 100$~pc in galaxies that tend to be the most massive
ellipticals.

\subsection{The Galactic Center}

The black hole at the Galactic center has a mass of $\sim 3 - 4\ts
10^6\:M_\odot$, appears to be embedded in hot gas (as seen in X-rays),
yet is underluminous by several orders of magnitude with respect to
thin disc accretion at the Bondi rate (Baganoff et al. 2003, and
references therein). Some, but probably not the majority, of the
diffuse X-ray emission on the arcsecond scales relevant to Bondi
accretion may be the dust-scattered halo of the X-ray emission from
inside the black hole's Bondi radius (Tan \& Draine 2004). Basic ADAF
models have been ruled out for the Galactic center due to polarization
measurements which indicate the density is even lower than that
predicted by ADAFs (Bower et al.  2003).

This system is probably more complicated than the supermassive black
holes in elliptical galaxies because the Bondi radius is only $\sim
0.1$~pc in size, which is relatively small compared to the feedback
scales associated with stellar wind bubbles and supernova remnants
from individual massive stars. For example, the dominant source of
mass in the region is likely to be from stellar winds from young,
massive stars. Some of these are even present inside the Bondi radius
(as estimated for the diffuse X-ray emitting gas) and because of the
relatively low mass of the black hole, these winds with speeds
$\sim$1000~km/s may be unbound and act to prevent diffuse gas from
accreting (Quataert 2004).

A further complication is that there are large amounts of neutral gas
in the vicinity (Herrnstein \& Ho 2004; Christopher et al. 2005).  The
sporadic infall of this material may well dominate the time-averaged
mass accretion rate.

\subsection{Quasars and AGN}

The issue of self-gravity and quasar accretion discs has been
discussed by Schlosman \& Begelman (1989), Goodman (2003) and Thompson
et al. (2005).  Our model herein has been developed with
low-luminosity galactic nuclei in mind, for which the central regions
are being fed at the Bondi rate from their host elliptical
galaxies. The accretion rates to the black holes powering quasars
approach the Eddington limited value: $\dot{M}_{d,{\rm E}} = 4\pi GM
l_{\rm E}/(\kappa_{e.s.} c \epsilon) \simeq 22 (l_{\rm
E}/\epsilon_{0.1}) M_9\:M_\odot\:{\rm yr^{-1}}$, where
$\kappa_{e.s}\simeq 0.4\:{\rm cm^2\:g^{-1}}$ is the electron
scattering opacity, $l_{\rm E}\equiv L/L_{\rm E}$ is the ratio of the
luminosity to the Eddington value, and $\epsilon\equiv L/(\dot{M}_d
c^2)\equiv 0.1 \epsilon_{0.1}$ is the radiative efficiency of the
accretion disc.

To have these accretion rates be the residual mass flux from a
star-formation supported disc on 100~pc scales requires star formation
rates that are $\sim 10^6$ times larger (eq.~\ref{mdotdiscthin}),
which are unrealistic. However, this estimate assumes the star
formation time is short compared to the radial advection time in the
disc. Thompson et al. (2005) show that the opposite may be true in
quasars, in which case the initial outer-disc mass feeding rate (or
equivalently the star formation rate) need only be factors of
$\sim$100 greater.

There may be additional complications, such as the supply of some very
low angular momentum material to the inner disc, which is not
gravitationally unstable.

%Rather, the material must be supplied at much smaller
%radii, where the disc becomes optically thick. 

%One natural ratio of
%star formation rate to black hole accretion rate to consider is the
%observed ratio of galactic spheroid stellar mass to black
%hole mass, $\sim 10^3$ (e.g. Magorrian et al. 1998).  Using
%eq.~(\ref{mdotdisc}) we see that to support a disc accretion rate to a
%$10^9\:M_\odot$ black hole of $1\:M_\odot\:{\rm yr^{-1}}$ with a star
%formation rate 1000 times larger, requires material to supplied to the
%disc at a radius of about 0.5~pc (using $\kappa_d=100\:{\rm
%  cm^{2}\:g^{-1}}$). Such mass supply rates could arise from clouds of
%mass $10^6\:M_\odot$ infalling and being disrupted on the orbital
%timescale at this radius. However, for such a model to be consistent
%with observations of black holes and their associated stellar
%populations, the stars would later have to migrate out to much larger
%($\sim$~kpc) distances after formation, which does not seem likely.
%One caveat associated with eq.~(\ref{mdotdisc}) is that it assumes
%only thermal pressure support in the disc, whereas, by comparison to
%the ISM of star-forming disc galaxies, magnetic and turbulent pressure
%may play a comparable role, thus increasing the accretion rate that
%can be supported by a given star formation rate.

The problem of the mass supply to quasars requires further work.  The
possibility remains that self-gravity in a disc is so severe that
black holes need to be supplied at much smaller radii, perhaps through
stellar disruption events (e.g. Hills 1975; Goodman 2003).

In lower luminosity AGN, such as Seyferts and LINERs, there is much
evidence for nuclear star formation (e.g.  Terlevich \& Melnick 1985).
These accretion discs are also expected to be self-gravitating, and
there is some observational evidence to support this (e.g. Kondratko,
Greenhill, \& Moran 2005). While the fueling of these nuclei may have
little to do with Bondi accretion (because of the presence of large
amounts of cold gas in an extended disc), the model of an accretion
disc supported by stellar feedback (\S4) probably is relevant,
predicting a general correlation of star formation and AGN activity.

%Morganti et al. (2001) pointed to a tentative correlation between the
%presence of HI absorption and starburst activity in more luminous
%radio galaxies.

%Show that overall contribution to elliptical colors is probably small.
%So I don't think this mechanism helps explain Fukugita's recent
%results. Check. Or general E+A phenomenon (Gunn).

%Trager et al.

\section{Conclusions}

Observations of the central regions of elliptical galaxies suggest
that supermassive black holes that reside there are fed at the Bondi
rate, of a few hundredths to a few tenths of a solar mass per year.
For standard thin disc accretion, these black holes are very
underluminous given this mass accretion rate. Previously, alternative accretion
models invoking inefficient radiation of gravitational energy, 
radially-dependent accretion rates, and outflows have been proposed
as solutions to this problem.  In this paper we have suggested an
additional process, likely occurring, which can be of comparable or
potentially greater importance: Bondi-fed star-forming
discs. Accretion occurs at the Bondi rate, but because the disc that
inevitably forms inside the Bondi radius is very gravitationally
unstable, most of the mass flux reaching it turns into stars.  Stellar
feedback is relatively weak so that only a small residual amount of
gas can be stabilized to reach the black hole at the center of the
disc. Our estimates of this residual mass flux are uncertain, but
appear to be potentially small enough that standard thin disc
accretion models can avoid the constraints imposed by the observed low
luminosities.  ADAFs, CDAFs, or other  models with outflows may still apply
in the very inner regions, but with a smaller outer boundary value for
their inward mass flux.

Our model is highly simplified: we mostly consider only a single
typical scale for the disc radius, whereas in reality gas will reach
the disc at a range of radii depending on the initial angular momentum
distribution; our quantitative estimates depend on the assumed stellar
IMF; we have adopted the standard alpha approximation for viscous
stresses in the disc; and we have mostly considered only thermal
pressure support in the disc, ignoring magnetic pressure.

As a result of these simplifications, it is vital to compare the model
as closely as possible to real systems, the best observed being
M87. 
%Ironically, M87 has a jet whose mechanical luminosity has been
%estimated to be $2 \times 10^{43}${\rm erg/sec} (Reynolds et al. 1996)
%which is $\sim 1/10$ of the Bondi rate. This would make it the least
%underfed of the known low luminosity ellipticals. Nevertheless...
We have described how the observational data on the nucleus of
this galaxy are consistent with the possibility that more than half 
of the Bondi 
accretion flow is depleted into stars: a thin disc of gas and dust,
apparently self-gravitating, is seen inside the Bondi radius and the
inferred star formation rate from the H$\alpha$ emission is consistent
with the Bondi accretion rate. This conclusion is not contradicted
even when current estimates of the mechanical luminosity of the M87 jet
are incorporated into the residual central engine accretion rates.  
Additional observations to detect molecular gas scales inside the Bondi radius 
are now feasible with the SMA. This
instrument is also capable of detecting continuum emission from dust
heated by the expected star formation.

\section*{acknowledgments}
We thank Henrik Beuther, George Field, Jeremy Goodman, Sebastian
Heinz, Luis Ho, Chris McKee, Vladimir Pariev, Linda Smith, and Sergei
Nayakshin for discussions and comments. JCT has been supported by a
Zwicky fellowship from the Inst. of Astronomy, ETH Z\"urich, by a
Spitzer-Cotsen fellowship from Princeton University and by NASA grant
NAG 5-10811. EGB acknowledges support from NSF grant AST-0406799, NASA
grant ATP04-0000-0016, and the KITP of UCSB, where this research was
supported in part by NSF Grant PHY-9907949.

\bsp

\label{lastpage}


\begin{thebibliography}{}

\bibitem[]{} Abel, T., Bryan, G.~L, \& Norman, M. L. 2002, Science, 295, 93

\bibitem[]{} Baath, L. B. et al. 1992, A\&A, 257, 31

\bibitem[Baganoff et al.(2003)]{2003ApJ...591..891B} Baganoff, F.~K.~et al.\ 2003, ApJ, 591, 891 

\bibitem[]{bh98} Balbus S.A. \& Hawley J.F., 1998, Rev Mod Physics, 72 1,
%Instability, turbulence, and enhanced transport in accretion discs.

\bibitem[Balbus \& Hawley(2002)]{2002ApJ...573..749B} Balbus, S.~A.~\& 
Hawley, J.~F.\ 2002, ApJ, 573, 749 


\bibitem[Begelman \& Chiueh(1988)]{bc88} Begelman, M.~C.~\& 
Chiueh, T.\ 1988, ApJ, 332, 872 

%\bibitem[]{binney03} Binney,~J., 2003, MNRAS, submitted (astro-ph/0308171)

\bibitem[]{} Biretta, J. A., Stern, C. P., \& Harris, D. E. 1991, AJ, 101, 1632


\bibitem[Bisnovatyi-Kogan \& Lovelace(1997)]{bkl97} 
Bisnovatyi-Kogan, G.~S.~\& Lovelace, R.~V.~E.\ 1997, ApJL, 486, L43 


%\bibitem[Bisnovatyi-Kogan \& Lovelace(2001)]{bkl01} 
%Bisnovatyi-Kogan, G.~S.~\& Lovelace, R.~V.~E.\ 2001, New Astronomy Review, 
%45, 663 


\bibitem[Blackman(1999)]{b99} Blackman, E.~G.\ 1999, MNRAS, 302, 723 

\bibitem[]{} Blandford, R. D., \& Begelman, M. C. 1999, MNRAS, 303, L1

\bibitem[]{} Bower, G. A., Heckman, T. M., Wilson, A. S., \& Richstone, D. O. 1997, ApJ, 483, L33

\bibitem[]{} Bower, G. C., Wright, M. C. H., Falcke, H., \& Backer, D. C. 2003, ApJ, 588, 331

\bibitem[]{} Carollo, C. M., Franx, M., Illingworth, G. D., \& Forbes, D. A. 1997, ApJ, 481, 710

\bibitem[]{} Chabrier, G. 2003, PASP, 115, 763

\bibitem[]{} Christopher, M. H., Scoville, N. Z., Stolovy, S. R., \& Yun, M. S. 2005, ApJ, 622, 346


\bibitem[Di Matteo et al.(2000)]{2000MNRAS.311..507D} Di Matteo, T., 
Quataert, E., Allen, S.~W., Narayan, R., \& Fabian, A.~C.\ 2000, MNRAS
311, 507 

\bibitem[]{} Di Matteo, T., Allen, S. W., Fabian, A. C., Wilson, A. S. \& Young, A. J. 2003, ApJ, 582, 133


\bibitem[]{} Fabian, A. C., \& Rees, M. J. 1995, MNRAS, 277, L55

\bibitem[]{} Faber, S. M., Tremaine, S, Ajhar, E. A. et al. 1997, AJ, 114, 1771

\bibitem[]{} Ford, H. C., Harms, R. J., Tsvetanov, Z. I. et al. 1994, ApJ, 435, L27


%\bibitem[Fisher, Kormendy, \& Bender(2003)]{2003AAS...20311616F} Fisher, 
%D.~B., Kormendy, J., \& Bender, R.\ 2003, American Astronomical Society 
%Meeting, 203,  
%Evidence From Surface Brightness Profiles for the Dissipative Merger Formation of Low-Luminosity Elliptical Galaxies


%\bibitem[Fukugita et al.(2004)]{2004ApJ...601L.127F} Fukugita, M., 
%Nakamura, O., Turner, E.~L., Helmboldt, J., \& Nichol, R.~C.\ 2004, ApJL, 
%601, L127 

\bibitem[]{} Gammie, C. F. 2001, ApJ, 553, 174

\bibitem[Georgakakis et al.(2001)]{2001MNRAS.326.1431G} Georgakakis, A., 
Hopkins, A.~M., Caulton, A., Wiklind, T., Terlevich, A.~I., \& Forbes, 
D.~A.\ 2001, MNRAS, 326, 1431 
%Cold gas in elliptical galaxies


\bibitem[Goodman (2003)]{g03} Goodman J., 2003, MNRAS, 339, 937. 


\bibitem[]{} Goodman, J.~\& Tan, J.~C.\ 2004, ApJ, 608, 108


%\bibitem[]{} Gruzinov A.V., 1998a, astro-ph/9809265. 

\bibitem[]{} Gruzinov, A.~V.\ 1998, ApJ, 501, 787 

\bibitem[]{} Harms, R. J., Ford, H. C., Tsvetanov, Z. I. et al. 1994, ApJ, 435, L35

\bibitem[]{} Herrnstein, R. M., \& Ho, P. T. P. 2004, ApJ, 620, 287

\bibitem[]{} Hills, J. G. 1975, Nature, 254, 295

\bibitem[]{} Ho, L. C. 1999, ApJ, 516, 672

\bibitem[]{ichimaru77} Ichimaru,~S., 1977, ApJ, 214, 840

\bibitem[]{} Jaffe, W., Ford, H. C., Ferrarese, L., van den Bosch, F., \& O'Connell, R. W. O. 1993, Nature, 364, 213

\bibitem[]{} Jaffe, W., Ford, H. C., O'Connell, R. W., van den Bosch, F. C., \& Ferrarese, L. 1994, AJ, 108, 1567

\bibitem[]{} Jaffe, W., \& McNamara, B. R. 1994, ApJ, 434, 110

\bibitem[]{} Jog, C. J., \& Solomon, P. M. 1984, ApJ, 276, 114

%\bibitem[]{jordi03} Miralda-Escud\'e J.\& Kollmeier J.A., 2003, submitted to ApJ, astro-ph/0310717

\bibitem[]{} Kennicutt, R. C. 1998, ApJ, 498, 541

\bibitem[]{} Kennicutt, R. C., Tamblyn, P., \& Congdon, C. W. 1994, ApJ, 435, 22

\bibitem[]{} Knapp, G. R., \& Rupen, M. P. 1996, ApJ, 460, 271

\bibitem[]{} Kondratko, P. T., Greenhill, L. J., \& Moran, J. M. 2005, ApJ, 618, 618

\bibitem[]{} Kormendy, J., \& Richstone, D. O. 1995, Ann. Rev. Astron. \& Astrophys. 33, 581

\bibitem[]{} Krajnovi\'c, D., \& Jaffe, W. 2004, A\&A, 428, 877

\bibitem[]{} Kroupa, P. 2002, Science, 295, 82

\bibitem[]{} Kuno, N., Nakai, N, Handa, T., \& Sofue, Y. 1995, PASJ, 47, 745

%\bibitem[]{landau88} Landau~L.D., Lifshitz~E.M., 1988,
%The Classical Theory of Fields. Nauka, Moscow

\bibitem[]{} Lauer, T. R., Faber, S. M., Gebhardt, K. et al. 2005, AJ, 129, 2138

\bibitem[]{} Leitherer, C. \& Heckman, T. M. 1995, ApJS, 96, 9

\bibitem[]{} Leitherer, C. et al. 1999, ApJS, 123, 3

\bibitem[]{} Li, A., \& Draine, B. T. 2001, ApJ, 554, 778

\bibitem[]{} Loewenstein, M., Mushotsky, R. F., Angelini, L., Arnaud, K. A. \& Quataert, E. 2001, ApJL, 555, L21

%\bibitem[]{} Magorrian, J., Tremaine, S., Richstone, D. et al. 1998, AJ, 115, 2285

\bibitem[]{} Macchetto, F., Marconi, A., Axon, D. J., Capetti, A., Sparks, W., \& Crane, P. ApJ, 489, 579

\bibitem[]{} Marshall, H. L., Miller, B. P., Davis, D. S., Perlman, E. S., Wise, M., Canizares, C. R., \& Harris, D. E. 2002, ApJ, 564, 683

\bibitem[]{} Martel, A. R., Ford, H. C., Bradley, L. D., Tran, H. D., Menanteau, F., Tsvetanov, Z. I., Illingworth, G. D., Hartig, G. F., \& Clampin, M. 2004, AJ, 128, 2758

\bibitem[]{} Martel, A. R., Turner, N. J., Sparks, W. B., \& Baum, S. A. 2000, ApJS, 130, 267

\bibitem[]{} Martin, C. L., \& Kennicutt, R. C. 2001, ApJ, 555, 301

\bibitem[Mathews \& Brighenti(2003)]{2003ARA&A..41..191M} Mathews, W.~G.~\& 
Brighenti, F.\ 2003, ARAA, 41, 191 
%Hot Gas in and around Elliptical Galaxies

\bibitem[]{} Miller, K. A., \& Stone, J. M. 2000, ApJ, 534, 398

\bibitem[]{} Muench, A. A., Lada, E. A., Lada, C. J., \& Alves, J. 2002, ApJ, 573, 366
  
\bibitem[]{} Narayan,~R., 2002, in Gilfanov,~M., Sunyaev,~R.,
  Churazov,~E., eds., Lighthouses of the Universe: The Most Luminous
  Celestial Objects and Their Use for Cosmology. European Southern
  Observatory, Garching, p.~405

\bibitem[]{narayan00} Narayan,~R., Igumenshchev,~I. V., \& Abramowicz,~M. A., 2000, ApJ, 539, 798

%\bibitem[]{narayan98} Narayan,~R. Mahadevan, R., Quataert E., 1998, in {\sl Theory of Black Hole Accretion Discs} eds. M. A. Abramowicz, G. Bjornsson, and J.E. Pringle. (Cambridge: CUP), p.148

\bibitem[]{} Narayan, R., Quataert, E., Igumenshchev, I.~V., \& Abramowicz, M.~A.\ 2002, ApJ, 577, 295 

\bibitem[]{} Narayan,~R. \& Yi,~I., 1995, ApJ, 452, 710

\bibitem[]{} Nayakshin, S. 2004, MNRAS, 352, 1028


\bibitem[]{} Pariev,~V., Igumenshchev,~I. V., \& Abramowicz,~M. A., 2000, ApJ, 539, 798



\bibitem[Pariev \& Blackman(2003)]{pb03} Pariev, V.~I.~\& Blackman, E.~G.\ 2003, astro-ph/0310167.


\bibitem[]{} Perlman, E. S., Sparks, W. B., Radomski, J., Packham, C., Fisher, R. S., Pi\~na, R., \& Biretta, J. A. 2001a, ApJ, 561, L51

\bibitem[]{} Perlman, E. S., Biretta, J. A., Sparks, W. B., Macchetto, F. D., \& Leahy, J. P. 2001b, ApJ, 551, 206



\bibitem[Quataert(1998)]{Q98} Quataert, E.\ 1998, ApJ, 500, 978 

\bibitem[]{quataert03} Quataert,~E., 2004, ApJ, 613, 322

\bibitem[]{} Quataert, E. \& Gruzinov, A. 1999, ApJ, 520, 248

\bibitem[]{} Quataert, E. \& Gruzinov, A. 2000, ApJ, 539, 809

\bibitem[]{} Quataert, E. \& Narayan, R. 2000, ApJ, 528, 236

\bibitem[]{} Randall, S. W., Sarazin, C. L. \& Irwin, J. A. 2004, ApJ, 600, 729

\bibitem[]{} Reynolds, C. S., Di Matteo, T., Fabian, A. C., Hwang, U., \& Canizares, C. R. 1996a, MNRAS, 283, L111

\bibitem[]{} Reynolds, C. S., Fabian, A. C., Celotti, A., \& Rees, M. J. 1996b, MNRAS, 283, 873

\bibitem[]{} Salpeter, E. E. 1955, ApJ, 121, 161

\bibitem[]{} Sarazin, C. L., Irwin, J. A. \& Bregman, J. N. 2001, ApJ, 556, 533

\bibitem[]{} Schaerer, D., Guseva, N. G., Izotov, Y. I., Thuan, T. X. 2000, A\&A, 362, 53

\bibitem[]{} Schlosman, I., \& Begelman, M. C. 1989, ApJ, 341, 685

\bibitem[]{} Sirko, E., \& Goodman, J. 2003, MNRAS, 341, 501

\bibitem[]{} Smith, L., J., Norris, R. P. F., Crowther, P. A. 2002, MNRAS, 337, 1309

\bibitem[]{} Spencer, R. E., \& Junor, W. 1986, Nature, 321, 753

%\bibitem[]{} Spitzer L., 1956, {\it Physics of fully Ionized Gases}, (New York: Wiley).

\bibitem[]{} Shakura N. I., \& Sunyaev, R. A. 1973, A\&A, 24, 337

\bibitem[]{} Shu, F. H. 1992, The Physics of Astrophysics, Vol. 2 (Mill Valley: Univ. Science Books)

\bibitem[]{} Sutherland, R. S. \& Dopita, M. A. 1993, ApJS, 88, 253

\bibitem[]{} Tan, J. C. 2000, ApJ, 536, 173

\bibitem[]{} Tan, J. C., \& Draine, B. T. 2004, ApJ, 606, 296

\bibitem[]{} Tan, J. C., \& McKee, C. F. 2004, ApJ, 603, 383

\bibitem[]{} Terebey, S., Shu, F. H., \& Cassen, P. 1984, ApJ, 286, 529

\bibitem[]{} Terlevich, R., \& Melnick, J. 1985, MNRAS, 213, 841

\bibitem[]{} Thompson, T. A., Quataert, E., \& Murray, N. 2005, ApJ, in press, (astro-ph/0503027)

\bibitem[]{} Tonry, J. L., Dressler, A., Blakeslee, J. P., Ajhar, E. A., Fletcher, A. B., Luppino, G. A., Metzger, M. R., \& Moore, C. B. 2001, ApJ, 546, 681

\bibitem[]{} Toomre, A. 1964, ApJ, 139, 1217

\bibitem[]{} Tomita, A., Aoki, K., Watanabe, M., Takata, T., \& Ichikawa, S.-I. 2000, AJ, 120, 123 

\bibitem[]{} van Bemmel, I. M., Dullemond, C. P., Chiaberge, M., \& Macchetto, F. D. 2004, astro-ph/0412374

\bibitem[]{} van den Bosch, F. C., Jaffe, W., \& van der Marel, R. P. 1998, MNRAS, 293, 343

\bibitem[]{} van Dokkum, P. G., \& Franx, M. 1995, AJ, 110, 2027

\bibitem[]{} Vila-Vilar\'o, B., Cepa, J., \& Butner, H. M. 2003, ApJ, 594, 232

\bibitem[]{} Waters, C. Z., \& Zepf, S. E. 2005, ApJ, 624, 656

\bibitem[Young(2002)]{2002AJ....124..788Y} Young, L.~M.\ 2002, AJ, 124, 788 
%Molecular Gas in Elliptical Galaxies: Distribution and Kinematics


\end{thebibliography}
\end{document}